# A First-principles Computational Framework for Quantum Decoherence in Complex Diamond Spin Environments

**Authors**

Huijin Park[1,2,3], Ha-young Jeong[4,5], Hyeonsu Kim[1], Christoph Findler[6], Fedor Jelezko[6], Sangwon Oh[2,5*], Junghyun Lee[4,7*], Giulia Galli[3,8,9], and Hosung Seo[1,2,4,10*]

*Correspondence to: seo.hosung@skku.edu; sangwonoh@ajou.ac.kr; jh_lee@kist.re.kr

**Affiliations**

[1]SKKU Advanced Institute of Nanotechnology, Sungkyunkwan University, Suwon, 16419, Korea

[2]Department of Physics and Department of Energy Systems Research, Ajou University, Suwon, Gyeonggi 16499, Republic of Korea

[3]Pritzker School of Molecular Engineering, University of Chicago, Chicago, IL, USA

[4]Center for Quantum Information, Korea Institute of Science and Technology, Seoul 02792, Republic of Korea

[5]Korean Research Institute of Standards and Science, Daejeon 34113, Republic of Korea

[6]Institute for Quantum Optics and Center for Integrated Quantum Science and Technology, Ulm University, Albert-Einstein-Allee 11, 89081 Ulm, Germany

[7]Division of Nano and Information Technology, KIST School, Korea University of Science and Technology, Seoul, Korea

[8]Center for Molecular Engineering and Materials Science Division, Argonne National Laboratory, IL, USA

[9]Department of Chemistry, University of Chicago, Chicago, IL 60637, USA

[10]Department of Quantum Information Engineering, Sungkyunkwan University, Suwon, 16419, Korea

## Abstract

Quantum decoherence induced by defects remains a major limitation for solid-state quantum technologies, yet predicting decoherence in realistic materials environments remains a formidable computational challenge. Consequently, complex defect populations are often approximated as homogeneous spin baths, obscuring the role of defect-specific electronic structure and spin dynamics. Here, we develop a predictive computational framework for quantum decoherence in complex diamond spin environments by combining first-principles electronic-structure calculations, quantum many-body spin-bath simulations, and experimental validation. The framework incorporates defect-resolved spin Hamiltonians derived from density functional theory and explicitly captures heterogeneous spin baths containing multiple paramagnetic defect species. Using diamond nitrogen-vacancy ensembles as a model quantum material platform, we investigate decoherence arising from mixed nitrogen-, vacancy-, and hydrogen-related defect environments. We show that quantum decoherence is governed not only by defect density but also by defect identity and bath composition, whose distinct electronic structures, hyperfine interactions, and spin dynamics produce qualitatively different coherence behavior. The calculations reveal that heterogeneous defect populations can either suppress or enhance decoherence, producing coherence trends that cannot be explained by conventional homogeneous-bath models. Magnetic-field-dependent Hahn-echo measurements performed on diamond samples with different defect concentrations quantitatively validate the framework. The calculations reproduce the observed coherence times and stretched-exponential decay behavior across a broad magnetic-field range and identify vacancy-related defects as critical contributors to decoherence beyond the conventionally assumed P1 spin bath. By linking atomistic defect properties to experimentally observable quantum coherence, our work establishes a predictive route for identifying hidden defect environments and provides a general computational strategy for understanding and optimizing decoherence in defect-based quantum materials.

## INTRODUCTION

The negatively charged nitrogen vacancy center ($NV^-$) in diamond has been extensively investigated as a primary platform for various quantum applications such as quantum sensing[1-6], quantum networks[7-9], and quantum computation[10-12]. Notably, NV-centers have demonstrated exceptional magnetic sensitivity[6, 13, 14] under a wide range of conditions, including ambient environments[15, 16], high pressures[17], and extreme magnetic fields[18-20], establishing NV-quantum sensing as one of the most versatile solid-state quantum sensing technologies. Leveraging these capabilities, NV spin-ensemble sensors have been widely adopted across diverse disciplines, spanning biomedical science[4, 21, 22] and condensed-matter physics [20, 23, 24]. In biological contexts, NV centers have enabled nanoscale detection of proteins[21] and single-neuron activities[22], while in solid-state systems, they have been employed to probe magnon dynamics[23] and image the Meissner effect in superconductors[24]. To further advance their practical impact, ongoing efforts[3, 5, 6, 25] focus on improving sensitivity, spatial resolution, and device integration for next-generation NV-based quantum sensors.

Optimizing the performance of NV-based quantum sensors requires maximizing two key metrics, which are achieving a high NV center density and extending the spin coherence time ($T_2$)[3, 25-27]. However, in diamond samples with high NV densities, $T_2$ is often limited by the presence of paramagnetic (PM) defects[28, 29]. These defects are unintentionally formed during NV center fabrication, primarily due to the low conversion efficiency from substitutional nitrogen ($N_s^0$) to $NV^-$, typically about 20% ($E_{conv} = \frac{[NV^-]}{[N_s^0]_i} \approx 20\ \%$)[28, 30-33], where $[N_s^0]_i$ denotes the initial nitrogen concentration introduced into the diamond host. As a result, a substantial fraction of nitrogen impurities remains in PM forms. PM defects possess unpaired electron spins ($S \geq 1/2$) that generate strong fluctuating magnetic fields, leading to significant shortening of the $T_2$ of nearby NV centers. Among them, the P1 center ($N_s^0$, $S$ =1/2) is the most prevalent PM defect species, forming with a conversion ratio, relative to the initial nitrogen concentration, of $[\mathrm{P1}]/[N_s^0]_i = 0.50 - 0.99$[3, 32-37]. Due to its high abundance, the P1 spin bath has been extensively investigated, both theoretically[38-42] and experimentally[27, 38, 43-48] as a dominant source of NV spin decoherence. In addition to P1 centers, other PM defects arising from nitrogen complexes, vacancies, and precursor gases (e.g., $CH_4/H_2$ mixtures) are formed concurrently during diamond growth and processing[33, 49]. However, their individual and combined contributions to NV spin decoherence remain largely unexplored. To optimize sensor performance, it is therefore important to establish bath-engineering guidelines supported by microscopic theoretical analysis, so that the defect environment can be understood and controlled to preserve long spin coherence at high NV densities. More broadly, achieving predictive control of quantum coherence requires computational frameworks capable of connecting atomistic defect properties to decoherence in realistic materials environments.

Recent experimental studies have begun to disentangle the distinct contributions of PM defects[3, 28, 37, 44, 50-54] to NV spin decoherence, which can be categorized according to their formation mechanisms. One major class consists of *N-related defects* such as $NV^0$ and $NVH^-$, which form with conversion ratios of $[NV^0]/[N_s^0]_i$ = 0.001 – 0.24, and $[NVH^-]/[N_s^0]_i$ = 0.04 – 0.26, respectively[3, 32-37]. Recent experimental work by Shinei *et al.*,[28] suggested that P1 and $NV^0$ centers are the main decoherence sources, whereas $NVH^-$ defects have comparatively minor effects. Another important class of PM defects comprises so-called *dark extra defects* (often referred to as X spins), which are not correlated with $[N_s]_i$, and are typically generated during irradiation and annealing processes[3, 51, 55-57]. These include vacancy- or hydrogen-related defects (*V/H-related defects*) such as $VH^-$, $VH^0$, $V^-$, $V^+$, vacancy clusters, and their complexes[3]. Wong *et al.*,[51] showed that vacancy complexes are a significant source of NV decoherence, which can be mitigated by high temperature annealing at approximately 1700 °C. Wang *et al.*,[52] further demonstrated that the charge states of vacancy complexes play a crucial role, as they determine the associated spin states. Moreover, they reported that the influence of PM-defect concentration on the NV coherence time ($T_2$) is dependent on whether the defect is a simple vacancy ($V^-$), or a weakly hyperfine-coupled spin such as $VH_2^-$, or $NVH^-$ center.

While experimental evidence for defect-dependent contributions to NV decoherence has recently emerged, a predictive theoretical framework capable of treating realistic heterogeneous defect environments is still lacking. To date, *N-related defects* and *V/H-related defects* have typically been treated as a semiclassical spin bath[38, 48, 58, 59] under the assumption that their spin-noise characteristics are similar to those of the P1 center. In many theoretical models, the presence of PM defects has also been represented as a bath of bare electron spins, independent of their specific defect identities[39, 60]. Such treatments implicitly neglect the internal degrees of freedom of PM defects, most notably hyperfine-coupled nuclear spins and defect-specific electronic structures, which can substantially modify their spin dynamics and associated magnetic noise spectra. In practice, distinct defect species within the spin bath must be treated as a mixed-spin many-body quantum system to accurately capture the underlying quantum dynamics. As interest in diamond bath engineering continues to grow, it has become increasingly important to develop a defect-resolved theoretical framework capable of quantitatively capturing these species-dependent contributions to NV decoherence. Such a framework would provide a route to predicting quantum decoherence directly from microscopic defect properties rather than relying on phenomenological spin-bath descriptions.

In this work, we develop a predictive computational framework for quantum decoherence in complex diamond spin environments by combining first-principles calculations, quantum many-body spin-bath simulations, and experiments. Specifically, we consider the contributions of multiple PM defects, including $NV^0$, $NVH^-$, $N_2^+$, $VH^0$, $VH^-$, $V^-$, and $V^+$ defects[3] by integrating the cluster-correlation expansion (CCE) method[40, 41, 61-67] with density functional theory (DFT). To systematically investigate the various contributions, we categorize the defects into *V/H related* or *N-related* classes and theoretically examine NV decoherence in both homogeneous (single-species) and heterogeneous (mixed-species) spin baths. We further compare the theoretical predictions with Hahn-echo measurements performed on diamond samples with different initial nitrogen concentrations under external magnetic fields. Through this combined approach, we establish a predictive connection between atomistic defect properties and experimentally observed quantum coherence, providing a general framework for understanding decoherence in defect-based quantum materials.

## RESULTS

### Spin properties of the paramagnetic defects

Figure 1 schematically illustrates the central spin model to compute the decoherence dynamics of an NV center interacting with a spin bath environment composed of various types of PM defects. The primary defects considered in this study are deep-level defects known to form in diamond, as well as their complexes, such as N, NV, NVH, $N_2$, VH, and V. Due to their deep-level nature, these defects can exist in multiple charge states, and their electronic and spin properties heavily depend on the specific charge state. Since here we focus on PM defects, we consider only charge states with non-zero electronic spin. Based on prior studies[3], we selected $N^0$ (P1), $NV^0$, $NVH^-$, $N_2^+$, $VH^0$, $VH^-$, $V^-$, and $V^+$ as representative PM defect species. Among them, P1, $NV^0$, $NVH^-$, $N_2^+$, $VH^0$, and $V^+$ are spin-1/2 centers, whereas $VH^-$ and $V^-$ exhibit spin quantum numbers of 1 and 3/2, respectively.

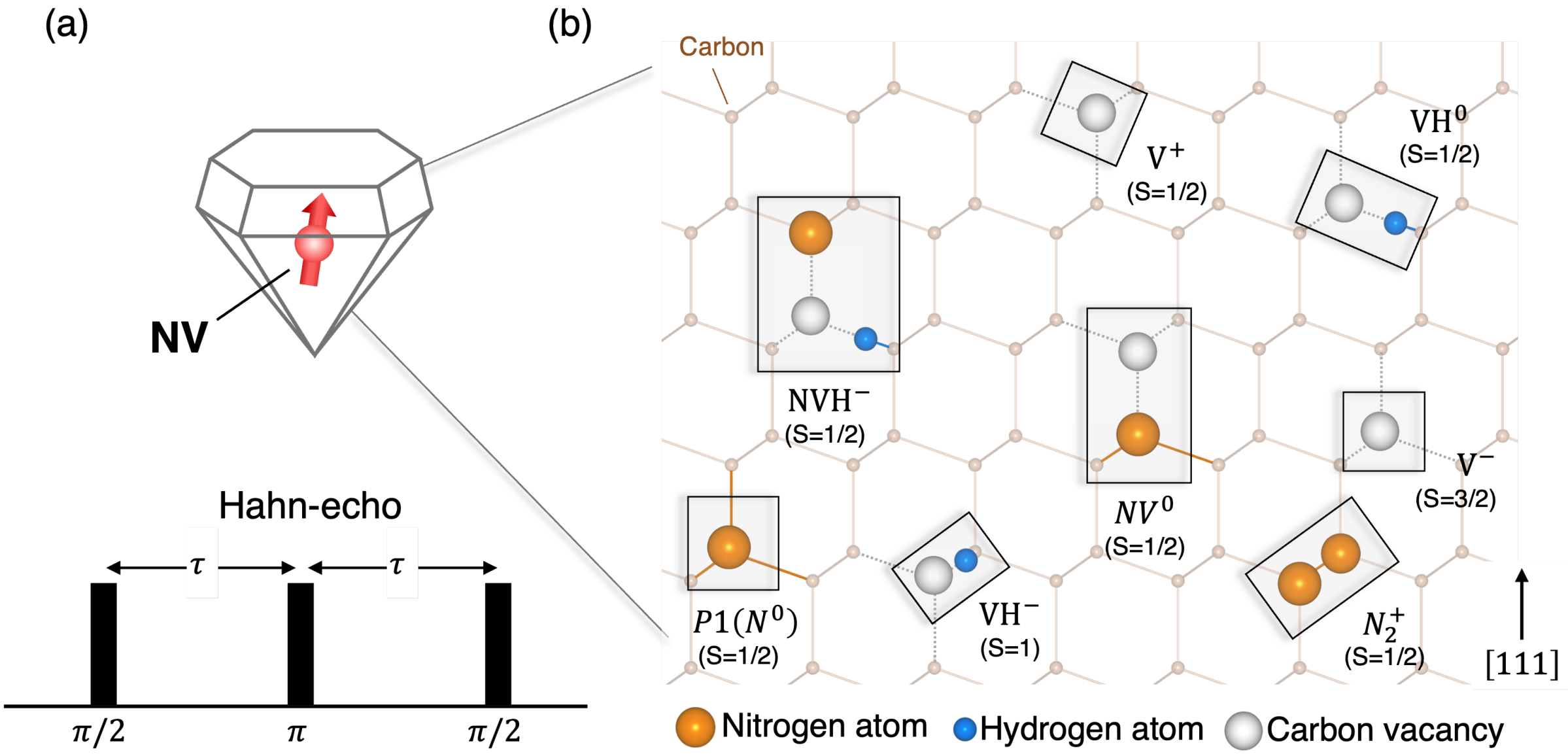

**Figure 1. Paramagnetic defects in diamond and central spin model.** (a) Schematic of the negatively charged nitrogen-vacancy in diamond subject to a Hahn-echo pulse sequence. (b) Representative paramagnetic defects considered in this study. Brown, blue, white, and orange spheres denote carbon, hydrogen, vacancies, and nitrogen atoms, respectively. An external magnetic field is applied along the [111] crystallographic axis.

We first discuss the geometrical degrees of freedom associated with each defect model, as they determine the electronic structure and, consequently, the spin Hamiltonian parameters. Two key structural factors play dominant roles: (i) Jahn–Teller (JT) distortions and (ii) the relative positions of impurity atoms within a defect complex. Both factors define the symmetry axis and local spin environment of each defect. For example, the P1 center exhibits four degenerate ground-state configurations arising from a static JT distortion, in which the nitrogen atom is displaced off-site from an ideal carbon lattice position along one of four crystallographic directions:[111], $[1\bar{1}\bar{1}]$, $[\bar{1}1\bar{1}]$, and $[\bar{1}\bar{1}1]$. In the case of the $NV^0$ defect[68], the nitrogen atom substitutes a carbon site adjacent to a vacancy (Figure 1b). The nitrogen atom can occupy four distinct positions relative to the vacancy, with each configuration defining a unique orientation of the defect's symmetry axis.

Similarly, for the $N_2^+$ defect, the two nitrogen atoms can be in four distinct configurations determining the direction of the symmetry axis. The $NVH^-$ defect exhibits a more complex structural landscape due to the presence of two species, N and H atoms (see Figure 1b). For a given nitrogen site, hydrogen can occupy three possible positions, each corresponding to a degenerate ground-state structure. Considering all combinations of N and H placements, we identify twelve distinct ground-state configurations with the same energy. For the $VH$ defect, the relative position of the hydrogen atom also gives rise to four possible orientations of the defect symmetry axis.

In all decoherence simulations, the PM defect's configuration within the bath is randomly sampled from its set of degenerate ground-state geometries (see Method). For each defect species, DFT calculations are performed to determine the optimized ground-state geometry, electronic structure (Supplementary Note 1.1) and spin Hamiltonian parameters (Supplementary Note 1.2 and 1.3). The defect-level diagrams and spin densities of these complexes are shown in Supplementary Figure 1. All defects considered here introduce deep electronic states within the diamond band gap, leading to highly localized spin centers.

We summarize the calculated hyperfine interaction parameters of the PM defects in Table 1. Notably, the hyperfine coupling strength varies substantially among different defect types, even for species sharing the same electron spin quantum number. For the $NVH^-$, $NV^0$, and $N_2^+$ defects (all with S=1/2), the computed parallel hyperfine interactions ($A_{\parallel}$) between the electron spin and the $^{14}$N nuclear spin are $-1.65$ MHz, $7.86$ MHz, and $128.70$ MHz, respectively. In addition, the $NVH^-$ defect exhibits an extra hyperfine interaction between the electron spin and the $^{1}$H nuclear spin, with $A_{\parallel} = -16.9$ MHz. For the $VH^0$ and $VH^-$ defects, which share the same structural motif (a V-H complex) but differ in charge state, the calculated $A_{\parallel}$ values differ by approximately 13%, highlighting the sensitivity of spin Hamiltonian parameters to defect charge state.

**Table 1. Calculated hyperfine interactions of paramagnetic defects in diamond.** Hyperfine parameters computed for each PM defect along its principal axis are compared with previously reported values in the literature. Experimental measurements are marked with an asterisk (*), while theoretical values from prior studies are denoted by a dagger (†).

| Defect | Atom | This study | | Reference | |
|---|---|---|---|---|---|
| | | $A_{\parallel}$ (MHz) | $A_{\perp}$ (MHz) | $A_{\parallel}$ (MHz) | $A_{\perp}$ (MHz) |
| $NVH^-$ | $^{14}N$ | $-1.65$ | $-2.11$ | - | - |
| | $^{15}N$ | $2.38$ | $2.90$ | $\pm 2.94$* [69] | $\pm 3.10$ |
| | $^{1}H$ | $-16.90$ | $5.90$ | $13.69$* [69] | $-9.05$ |
| $NV^0$ | $^{14}N$ | $7.86$ | $3.92$ | $11.19$† [68] | $6.74$ |
| | $^{15}N$ | $-11.03$ | $-5.5$ | $-15.69$† [68] | $-9.445$ |
| $N_2^+$ | $^{14}N$ | $128.70$ | $54.84$ | $159.78$† [70] | $85.54$ |
| | $^{15}N$ | $-180.53$ | $-76.93$ | - | - |
| $VH^0$ | $^{1}H$ | $23.09$ | $-13.78$ | $27$* [71] | $-6$ |
| $VH^-$ | $^{1}H$ | $-1.92$ | $-0.45$ | $1.10$* [72, 73] | $1.95$ |

The calculated hyperfine parameters are broadly consistent with previous theoretical and experimental reports in terms of their overall magnitude and anisotropy. The remaining differences can be attributed to the strong sensitivity of hyperfine couplings to computational details and strain effects that are not included in our calculations. In particular, the accuracy of the Fermi-contact term can be improved by describing the core region more accurately, and previous work has shown that this is important for reducing the residual error in DFT hyperfine calculations[74]. For $VH^-$ and $NVH^-$, the opposite sign of the $^{1}H$ hyperfine parameters likely arises from the experimental sign convention, which was assigned based on a dipolar model rather than directly determined through observation[69].

We further evaluate the nuclear quadrupole interaction by computing the electric field gradient

(EFG) at each nuclear site, as summarized in Table 2. The quadrupole term appears only for nuclei with $I > \frac{1}{2}$, such as $^{14}$N. The largest quadrupole coupling is found for $NVH^-$, with $P_\parallel = -5.07$ MHz, in good agreement with experimental measurements[69]. For $NV^0$ and $N_2^+$, $P_\parallel$ values are $-4.84$ and $-2.48$ MHz, respectively.

**Table 2. Calculated nuclear quadrupole interactions of paramagnetic defects in diamond**. The computed quadrupole parameters for each defect along its principal axis are compared with previously reported literature values. The parallel and perpendicular components are defined as $P_\parallel = 3Q_{zz}/2$ and $P_\perp = (Q_{xx} + Q_{yy})/2$, respectively. Experimental measurements are marked with an asterisk (*), and theoretical values from prior studies are denoted by a dagger (†).

| Defect | Atom | This study | | Reference | |
|---|---|---|---|---|---|
| | | $P_\parallel$ (MHz) | $P_\perp$ (MHz) | $P_\parallel$ (MHz) | $P_\perp$ (MHz) |
| $NVH^-$ | $^{14}$N | $-5.07$ | 1.689 | $-4.8$* [69] | - |
| $NV^0$ | $^{14}$N | $-4.84$ | 1.61 | - | - |
| $N_2^+$ | $^{14}$N | $-2.48$ | 0.83 | $-2.45$† [70] | - |

We note that PM defects with electronic spin quantum numbers larger than S=1/2 exhibit zero-field splitting (ZFS). For $VH^-$, our calculations yield ZFS parameters of $D = 2776.61$ MHz and $E = 10.03$ MHz, in good agreement with previous experimental measurements ($D = 2706$ MHz)[72]. The ZFS has not been experimentally resolved for the $V^-$ defect, likely due to its $T_d$ symmetry, which renders the splitting negligibly small[75]. Our calculations give $D = 52$ MHz and $E = 10$ MHz for $V^-$, both of which are negligibly small within the uncertainty of our DFT calculations. Hence, we consider the ZFS of $V^-$ as negligible and exclude it from the decoherence simulations.

**Impact of paramagnetic defects on NV decoherence**

Before addressing mixed-bath-driven decoherence, we first examine homogeneous spin-1/2

defect bath models to clarify the role of defect type in NV decoherence. Figures 2a and 2b show the calculated coherence functions of the NV ensemble and the corresponding $T_2$ values, respectively, for all defect species at a concentration of 12 ppm. Here, the magnetic field was set to a moderately high value of 322.7 G to avoid cross-relaxation regime (~ 512 G). We observe that $T_2$ varies substantially depending on the specific PM defect species. The computed $T_2$ values for $NV^0$ and $NVH^-$ baths are 13.7 $\mu s$ and 19.7 $\mu s$, respectively, both significantly shorter than the 25.1 $\mu s$ obtained for the P1 bath (Figure 2b). The $V^+$ and $VH^0$ baths exhibit also shorter coherence times of 5.3 $\mu s$ and 13.7 $\mu s$, respectively. In contrast, NV coherence in an $N_2^+$ defect bath decays much more slowly than in the P1 bath (see Supplementary Figure 2). The relative magnitude of $T_2$ reflects the magnetic noise strength generated by each defect bath: a longer $T_2$ indicates weaker effective magnetic noise and more suppressed bath spin dynamics. These results demonstrate that even among spin-1/2 defects, decoherence is strongly defect-type dependent.

To gain microscopic insight into the defect-dependent behavior, we analyze the dynamics of a representative pair of PM defects. At sufficiently large magnetic fields, energy-conserving electron-spin flip-flop (FF) transitions between two defects dominate the bath dynamics. In addition, the electron and nuclear spin product state wavefunctions closely approximate the eigenstates of the defect spin Hamiltonian. As discussed previously for P1 pairs[39], the FF dynamics of a defect pair can be effectively described using a pseudo-spin model. As illustrated in Figure 2c, the model considers electron-spin FF transitions between states $\left|m_S^{(i)}, m_I^{(i)}; m_S^{(j)}, m_I^{(j)}\right\rangle$ and $\left|m_S^{(i)} \pm 1, m_I^{(i)}; m_S^{(j)} \mp 1, m_I^{(j)}\right\rangle$, where $m_S^{(i)}$ and $m_I^{(i)}$ denote the electron and nuclear spin projections of the $i^{\text{th}}$ PM defect. For example, these projections correspond to $|\uparrow\downarrow\rangle \leftrightarrow |\downarrow\uparrow\rangle$ for a $V^+$ pair, and $|\uparrow 0 \downarrow 1\rangle \leftrightarrow |\downarrow 0 \uparrow 1\rangle$ for a $NV^0$ pair. Within this framework, the FF transitions are governed by two competing factors: 1) the pseudo-spin transition rate ($\Omega_{FF}$), mainly induced by electron-electron dipolar coupling ($\mathcal{H}_{ij}$); 2) the energy detuning ($\Delta_{FF}$) arising from the intra-defect interactions of the individual defects ($\mathcal{H}_{PM1} + \mathcal{H}_{PM2}$). The balance between $\Omega_{FF}$ and $\Delta_{FF}$ determines the efficiency of FF processes and thus the magnetic noise spectrum experienced by the NV center (see Supplementary Note 3 for further details).

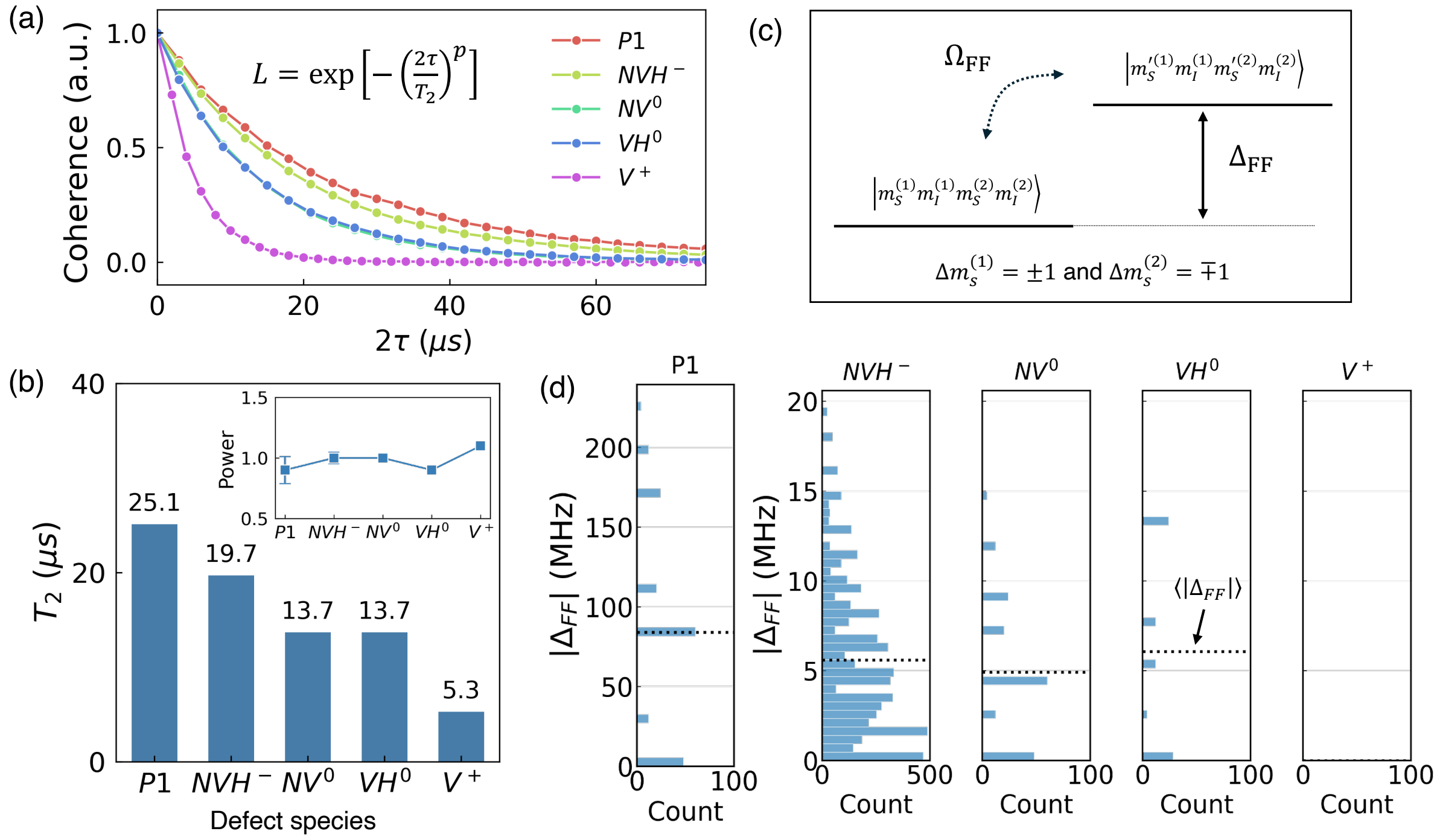

**Figure 2. NV decoherence in homogeneous PM defect baths.** (a) Computed coherence functions of NV centers in homogenous defect baths with a defect concentration of 12 ppm under a magnetic field of 322.7 G. (b) Extracted $\boldsymbol{T_2}$ and $\boldsymbol{p}$ (inset) parameters for each bath. They are obtained by fitting the coherence functions $L$ using a stretched exponential decay model. (c) Energy-level diagram illustrating electron-spin flip-flop transitions within a defect pair, where $\boldsymbol{\Delta m_s^{(i)} = m_s'^{(i)} - m_s^{(i)}}$. Here, $\boldsymbol{\Delta_{FF}}$ and $\boldsymbol{\Omega_{FF}}$ denote the energy detuning and the transition rate, respectively. (d) Calculated energy detuning $\boldsymbol{\Delta_{FF}}$ for all possible FF transitions within a homo-defect pair. For N-related defects, the $^{14}$N nuclear spin is considered in the evaluation of $\boldsymbol{\Delta_{FF}}$. $\boldsymbol{\Delta_{FF}}$ depends on both the defect species and the orientations of their symmetry axes, owing to variations in their HF interactions. The dashed lines denote average values, which is defined as $\langle|\boldsymbol{\Delta_{FF}}|\rangle = \frac{\mathbf{1}}{\boldsymbol{N}}\sum_{\boldsymbol{c}=\mathbf{1}}^{\boldsymbol{N}}\left|\boldsymbol{\Delta_{FF}^{(c)}}\right|$, where $\boldsymbol{N}$ is the total number of FF channels for a given homo-defect pair, and $\boldsymbol{c}$ indexes individual flip-flop channels.

We next compare the values of $\Omega_{FF}$ and $\Delta_{FF}$ across different homogeneous defect baths considered in Figure 2a to identify which factor primarily governs defect-type-dependent spin bath dynamics. The transition rate $\Omega_{FF}$ depends on the inter-defect distance and the electron spin quantum number (see Supplementary Equation S12). Since all defect species in Figure 2a have the same electron spin (S=1/2) and identical defect concentrations, the distribution of $\Omega_{FF}$ is essentially unchanged among the different homogeneous baths. Therefore, $\Omega_{FF}$ is not the primary factor responsible for the observed defect-type-dependence.

In contrast, the energy detuning $\Delta_{FF}$ exhibits strong defect-specific variation. In Figure 2d, we compute the distribution of $\Delta_{FF}$ for all possible FF transitions of a homo-defect pair $(i, j)$ across the possible state $\left|m_S^{(i)}, \left\{m_I^{(i,a)}\right\}; m_S^{(j)}, \left\{m_I^{(j,b)}\right\}\right\rangle$, which is given by $\Delta_{FF} = \pm \sum_a A_{zz,ia} m_I^{(i,a)} \mp \sum_b A_{zz,jb} m_I^{(j,b)}$, where $A_{zz,i}$ is the $zz$ components of the lab-frame hyperfine tensor for nuclear spin associated with the $i$-th defect; $m_I^{(i,a)}$ denote its nuclear spin projection. The indices $a$ and $b$ label individual nuclear spins associated with defects $i$ and $j$ (see Supplementary Note 3 for further details). We find that the variation of $T_2$ across different baths (Figure 2b) is well explained by the defect-dependent trends in $\Delta_{FF}$. Specifically, the averaged $\Delta_{FF}$ values are 83.91, 5.59, 6.06, 4.93, and 0.0 MHz for $P1$, $NVH^-$, $VH^0$, $NV^0$ and $V^+$ baths, respectively. For the P1 bath, the flip-flop channels are strongly suppressed by the largest $\Delta_{FF}$, which mainly originates from the JT distortion and the hyperfine interaction with the nitrogen nuclear spin, leading to the longest $T_2$ at the same defect concentration. For $NVH^-$ bath, the presence of the hydrogen atom introduces additional geometric configurations and nuclear spin states, increasing the probability of misaligned defect pairs and enhancing effective detuning. Consequently, the $T_2$ in the $NVH^-$ bath is significantly longer than in the $VH^0$ and $NV^0$ baths, even though their averaged $\Delta_{FF}$ values are comparable. These results show that hyperfine interactions, that are governed by the interplay between electronic structure and defect geometry, play a decisive role in determining NV decoherence in homogeneous defect baths.

**NV coherence dynamics in the mixed spin bath**

We now examine the decoherence dynamics of NV ensembles in mixed spin baths composed of dominant P1 centers and additional minor paramagnetic defects. Although P1-induced decoherence is weaker than that arising from other PM defects, P1 centers remain a dominant noise source in ensemble NV systems due to the high nitrogen densities introduced during growth. In the mixed-bath model, we include not only couplings between defects of the same species but also inter-species couplings between different defect species.

To systematically investigate the role of extra PM defects, we consider two distinct mixed-bath models, based on known common diamond synthesis and post-processing methods. The first model is *a defect-conversion model*, motivated by processes in which P1 centers are partially converted into other N-related defects such as $NVH^-$ or $NV^0$ during high-temperature annealing or related treatments. In this scenario, the total nitrogen concentration is conserved but redistributed among different defect species. In our simulations, as schematically illustrated in Figure 3a, we fix the total nitrogen concentration to the initial P1 concentration $[P1]_i$, while allowing for a fraction of P1 centers to convert into other N-related defects. This condition is expressed as: $[P1]_i = [P1]_f + [NX]$, where $[P1]_f$ denotes the remaining P1 concentration after conversion, and $[NX]$ represents the concentration of newly formed *N-related defects* (e.g. $NV^0$ or $NVH^-$).

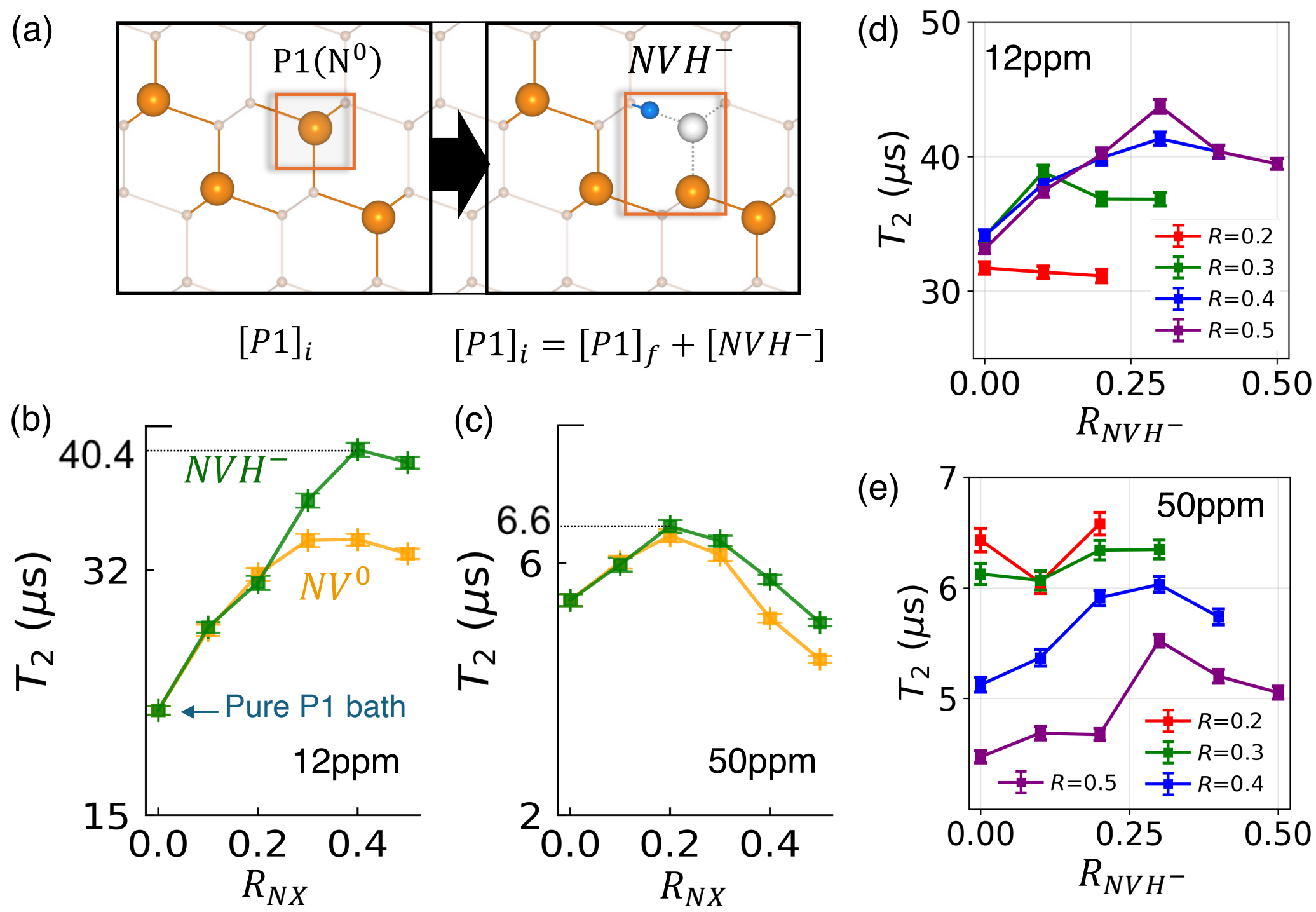

**Figure 3. Variation of the coherence time T₂ in the defect conversion model** (a) Schematic of the defect conversion model illustrating the transformation of P1 centers into other N-related defects (e.g. $NVH^{-}$). Orange, white and blue ball denote nitrogen atom, carbon vacancy and hydrogen atom respectively. The orange box indicates a site converted defect from P1 to $NVH^{-}$. (b, c) Computed $T_2$ values in two-species mixed bath as a function of the conversion ratio $R_{NX}$, for (b) $[P1]_i$ = 12 ppm and (c) $[P1]_i$ = 50 ppm. (d, e) $T_2$ values in three-species mixed bath ($P1$- $NV^0$-$NVH^{-}$) as a function of $R_{NVH^-}$ for (d) $[P1]_i$ = 12 ppm and (e) $[P1]_i$ = 50 ppm. The total conversion ratio $R = R_{NV^0} + R_{NVH^-}$ is varied from 0.2 to 0.5, while the individual fractions $R_{NV^0}$ and $R_{NVH^-}$ are independently swept within that range. The applied magnetic field is 322.7 G for 12 ppm and 325.7 G for 50 ppm.

Figures 3b and 3c present the computed $T_2$ values as a function of the defect conversion ratio $R_{NX} = \frac{[NX]}{[P1]_i}$ for fixed total nitrogen concentrations of $[P1]_i$ = 12 ppm and $[P1]_i$ = 50 ppm, respectively. Surprisingly, we find that $T_2$ exhibits a distinct, non-monotonic behavior depending on the conversion ratio $R_{NX}$. In both $P1$-$NVH^{-}$ and $P1$-$NV^0$ mixed baths, $T_2$ initially increases with increasing $R_{NX}$, reaching maximum values at $R_{NX}$ = 0.4 for $[P1]_i$ = 12 ppm and $R_{NX}$ = 0.2 for $[P1]_i$ = 50 ppm. As $R_{NX}$ approaches 1.0, the bath evolves toward a homogeneous *NX*-defect bath, and the resulting $T_2$ decreases significantly, becoming shorter than that of the P1-only bath (see Figure 2). For $[P1]_i$ = 12 ppm, the maximum $T_2$ values reach 40.4 $\mu s$ and 34.1 $\mu s$ for the $P1$-$NVH^-$ and $P1$-$NV^0$ mixed baths, respectively, which is almost 1.6 times and 1.4 times the $T_2$ of the P1-only bath. For $[P1]_i$ = 50 ppm, the corresponding maximum values are 6.58 $\mu s$ and 6.43 $\mu s$. Our results show that introducing $NVH^-$ or $NV^0$ as a minority species into a P1-dominant bath can substantially enhance NV coherence beyond the P1-only limit.

To elucidate the origin of the $T_2$ enhancement in mixed baths, we extend the pseudo-spin model developed for homogeneous defect pairs to heterogeneous defect pairs. Unlike homo-defect pairs, the energy detuning $\Delta_{FF}$ of hetero-defect pairs generally remains non-zero even when their symmetry axes are aligned. This intrinsic detuning arises from the difference in hyperfine interactions between distinct defect species. For instance, in an on-axis aligned $P1$-$NV^0$ pair, the transition $|\uparrow 1 \downarrow 1\rangle \leftrightarrow |\downarrow 1 \uparrow 1\rangle$ exhibits a substantial detuning given by $\Delta_{FF} = A_{zz,NV^0} - A_{zz,P1} = -106.2$ MHz. More generally, the calculated average detuning $\langle|\Delta_{FF}|\rangle$ for on-axis aligned hetero-defect pairs is 78.7 MHz for $P1$-$NVH^-$ and 77.7 MHz for $P1$-$NV^0$ pair. For hetero-defect pairs aligned along identical off-axes, the average detuning is also large, with values of 57.2 MHz and 57.7 MHz for $P1$-$NVH^-$and $P1$-$NV^0$ pair, respectively. Therefore, our results show that converting a fraction of P1 centers into other N-related PM defects increases the average energy detuning in the bath, suppresses FF transitions, and consequently prolongs the NV coherence.

Notably, however, the $T_2$ enhancement due to the compositional mixing is much more pronounced in the $[P1]_i$ = 12 ppm bath than in the $[P1]_i$ = 50 ppm bath. We find that the slope of $T_2$ as a function of $R_{NVH^-}$ in the regime where $T_2$ increases is 44.32 $\mu s$ for $[P1]_i$ = 12 ppm (Figure 3b) and 5.85 $\mu s$ for $[P1]_i$ = 50 ppm (Figure 3c). This difference can be understood within the pseudo-spin framework, where the decoherence amplitude of a defect pair can be expressed as $\kappa = \frac{\Omega_{FF}^2 \delta_{NV-PM}^2}{[\Omega_{FF}^2 + \Delta_{FF}(\Delta_{FF} + \delta_{NV-PM})]^2 + \Omega_{FF}^2 \delta_{NV-PM}^2}$ (See Supplementary Equation S22). In the denser mixed bath ($[P1]_i$ = 50 ppm), the average inter-defect distance is smaller, leading to a significantly larger dipolar coupling $\Omega_{FF}$ for defect pairs. As $\Omega_{FF}$ increases, the suppression effect of $\Delta_{FF}$ becomes less effective unless $\Delta_{FF} \gg \Omega_{FF}$. Consequently, in dense baths the FF probability is increasingly governed by $\Omega_{FF}$. Thus, while compositional mixing enhances detuning and suppresses FF dynamics, its effectiveness diminishes as the bath density increases, explaining why defect-induced $T_2$ enhancement is more efficient in the lower-concentration regime.

To further explore the enhancement mechanism described above, we compute $T_2$ for a three-species mixed bath ($P1$- $NV^0$-$NVH^-$) at each $[P1]_i$ as shown in Figure 3d, and 3e. The total conversion ratio $R = R_{NV^0} + R_{NVH^-}$ is varied from 0.2 to 0.5, while the individual fractions $R_{NV^0}$ and $R_{NVH^-}$ are independently swept within that range. For $[P1]_i = 12$ ppm, the $T_2$ reaches a maximum value of 43.7 $\mu s$ at $R_{NV^0} = 0.2$ and $R_{NVH^-} = 0.3$. At the higher density of $[P1]_i = 50$ ppm, the maximum $T_2$ is 6.6 $\mu s$, occurring in the absence of $NV^0$ defects ($R_{NV^0} =$ 0.0 and $R_{NVH^-} = 0.1$). Compared to the two-species mixed baths, the additional enhancement in the three-species case is modest. These results indicate that the primary driver of $T_2$ enhancement is the suppression of hetero-pair FF transitions arising from hyperfine mismatches, rather than the mere increase in the number of defect species. (See Supplementary Figure 4 for complete $T_2$ data across multiple-species mixed baths.)

The second mixed-bath model considered here is an *extra-defect model*, motivated by experimental processes such as electron irradiation, during which additional vacancy-related defects are generated. In this scenario, unlike in the defect-conversion model, the total nitrogen concentration remains fixed, while extra V/H-related paramagnetic defects are introduced into the bath. As schematically illustrated in Figure 4a, we fix the nitrogen concentration to the initial P1 density $[P1]_i$, and introduce additional V/H-related defects with concentration [X]. Consequently, the total paramagnetic defect density increases to $[P1]_i + [X]$, where $[X]$ denotes the concentration of the extra vacancy- or hydrogen-related defects.

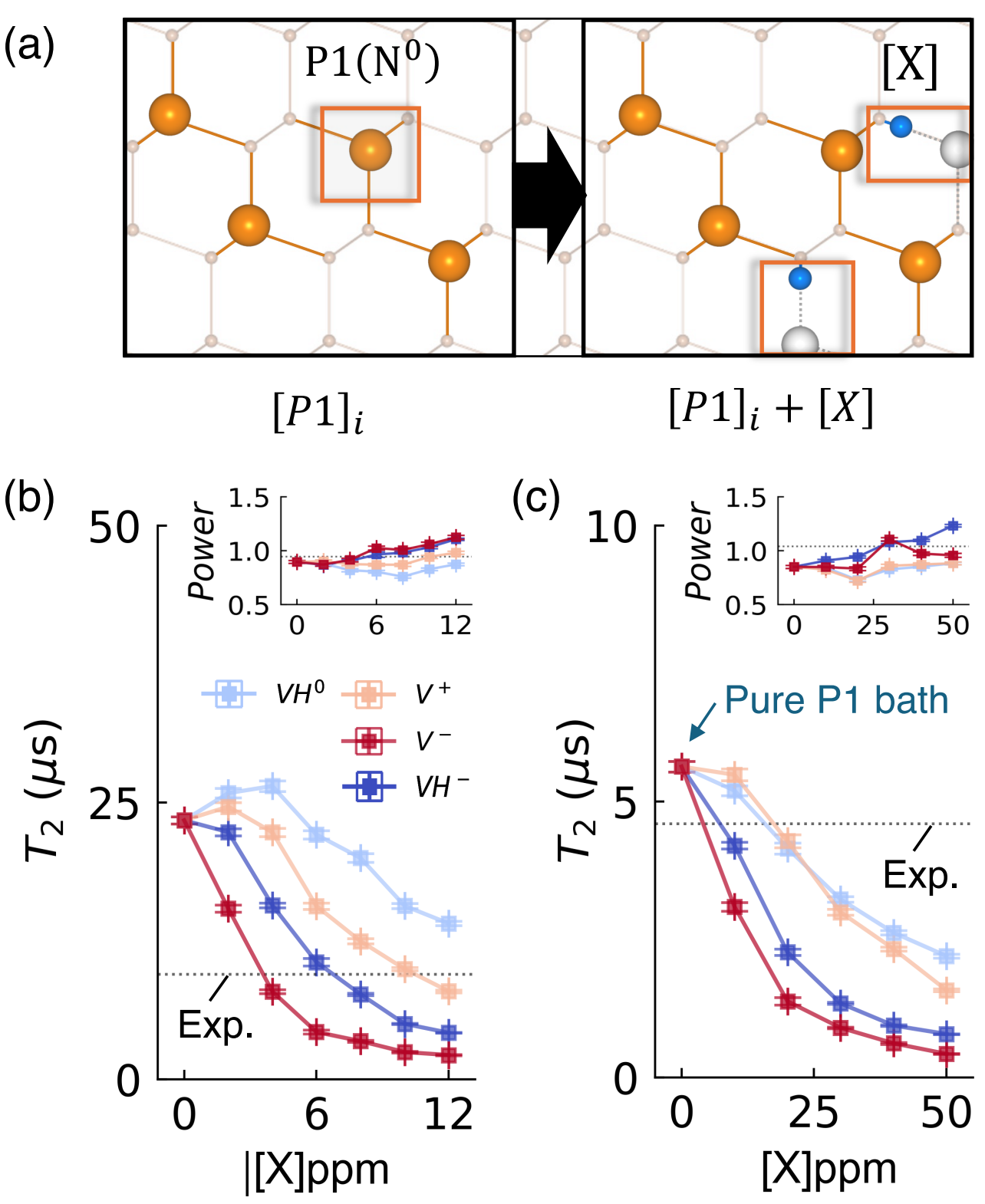

**Figure 4. Variation of the coherence time $T_2$ in extra defect model** (a) Schematic of the extra defect model. Orange boxes labeled $[X]$ represent V/H-related defects introduced into the bath. (b, c) Computed $T_2$ and $p$ values as a function of $[X]$ for (b) $[P1]_i$ = 12 ppm and (c) $[P1]_i$ = 50 ppm. Black dotted lines indicate experimentally extracted $T_2$ and $p$ values. Experimental details are described in a later section.

Figure 4b, and 4c show $T_2$ as a function of the V/H-related defect concentration [X], which is varied from 0 to $[P1]_i$. In contrast to the defect-conversion model, we observe that $T_2$ generally

decreases as [X] increases, reflecting the increase in the total number of electron spins in the bath. However, the rate of this decrease depends strongly on the specific defect species. At $[P1]_i$= 12 ppm (Figure 4b), the introduction of $V^-$ and $VH^-$ defects leads to a pronounced reduction in $T_2$, characterized by a steep decline even at small [X]. In contrast, for $V^+$- and $VH^0$-mixed baths, $T_2$ exhibits a weakly non-monotonic dependence on [X], with shallow maxima at [$VH^0$] = 4 ppm and [$V^+$] = 2 ppm, followed by a gradual decrease at higher concentrations. At the higher P1 density of $[P1]_i$ = 50 ppm (Figure 4c), $T_2$ decreases monotonically with increasing [X] for all defect species, and no enhancement is observed. The difference in the decreasing trends between $V^-$/$VH^-$-mixed baths and $V^+$/$VH^0$-mixed baths can be attributed to their distinct spin quantum numbers and associated dipolar interaction strengths. Specifically, defects such as $VH^-$ (with S = 1) and $V^-$ (with S = 3/2) generate stronger dipolar couplings than spin-1/2 defects such as $V^+$ and $VH^0$, leading to a more rapid decoherence. Overall, V/H-related defects lead to a reduction of $T_2$, and the extent of decoherence depends sensitively on their charge-state-dependent spin properties and corresponding dipolar interaction strengths.

**Experimental results and the presence of parasitic spins**

To investigate experimentally the presence of PM defects beyond P1 centers and their impact on NV decoherence, we perform Hahn-echo measurements of NV ensemble coherence as a function of the external magnetic field $B_0$ (20–330 G) for three $^{12}$C-enriched diamond samples with [$^{15}N_s^0$] = 10 ppm, [$^{14}N_s^0$] = 12 ppm and [$^{14}N_s^0$] = 50 ppm (Figure 5). In particular, the external magnetic field $B_0$ serves as a key tuning parameter for resolving decoherence in mixed spin baths. In each defect, the Zeeman term competes with other spin-interaction terms, such as hyperfine and zero-field-splitting interactions. This competition reshapes the energy-level structure, changes the degree of level mixing, and modifies the energy-matching conditions for spin transitions. As a result, varying $B_0$ can modify transition channels besides flip-flop processes, thereby affecting NV decoherence. Because different defect species respond differently to the applied field owing to their distinct spin interactions, we investigate their signatures in the magnetic-field dependence of NV decoherence.

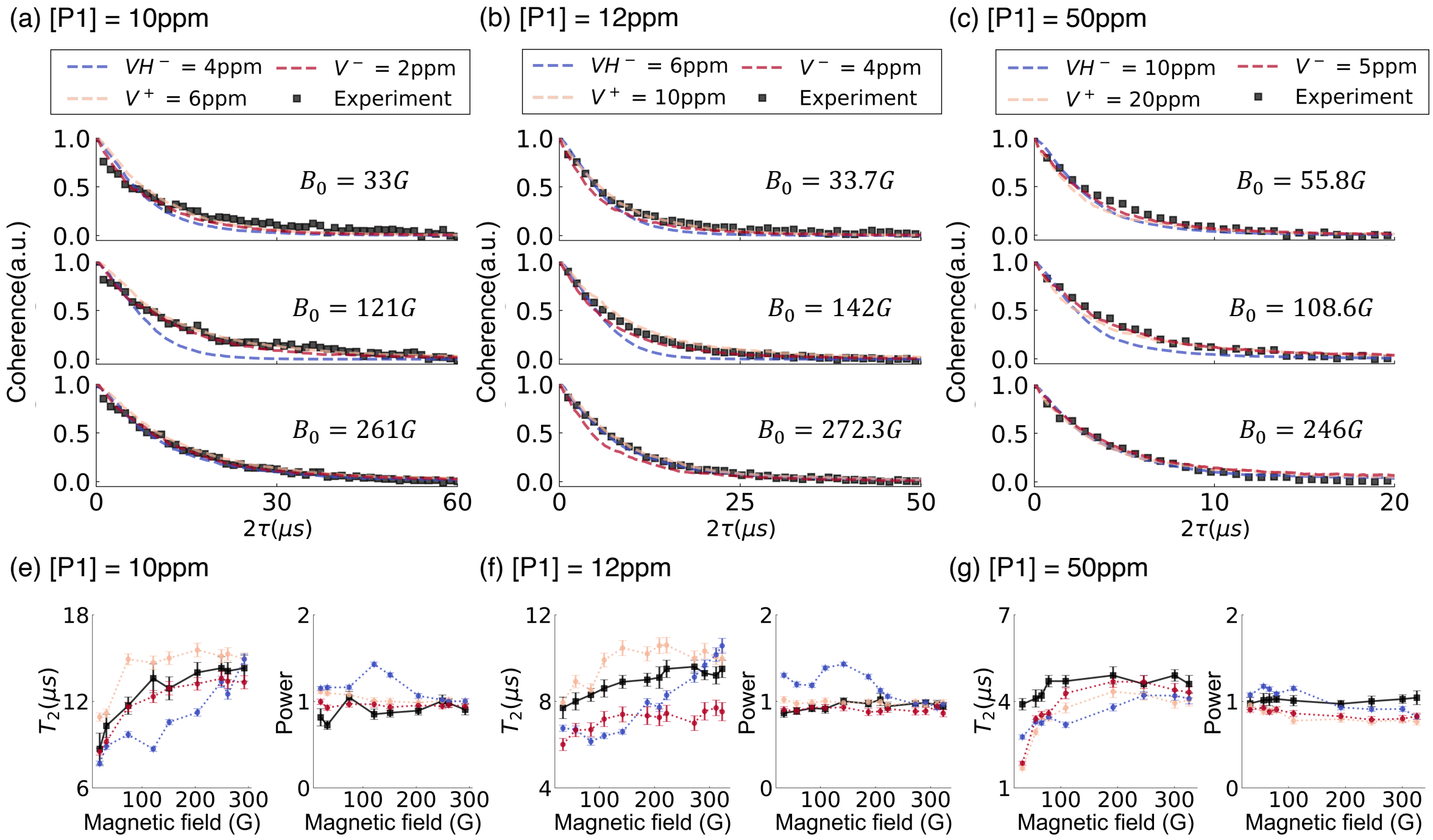

**Figure 5. Comparison between experiment and theory for Hahn-echo $T_2$ and $p$ values** (a-c) Measured NV spin ensemble coherence at three representative magnetic fields for initial P1 concentrations of 10 ppm, 12 ppm and 50 ppm, respectively. Black squares represent experimental data, and dashed lines indicate the results of CCE simulations. Red, light orange, blue dashed lines correspond to theoretical coherence in the mixed spin bath containing P1 centers together with $\boldsymbol{V^-}$, $\boldsymbol{V^+}$, and $\boldsymbol{VH^-}$ defects, respectively. (d-f) $\boldsymbol{T_2}$, and $\boldsymbol{p}$ values as a function of magnetic field ranging from 20 G to 330 G for the three samples.

Figure 5a-c display the measured coherence at three representative magnetic field strengths, while Figure 5d-f summarize the extracted $T_2$ and stretching exponent $p$ values as a function of $B_0$. For all three samples, both $T_2$ and $p$ values exhibit a consistent and gradual increase with increasing magnetic field. For example, in the 10-ppm sample, $T_2$ increases from 8.7 $\mu s$ at 20.5 G to 14.3 $\mu s$ at 292 G, while $p$ rises from 0.8 to 0.9. The 12-ppm sample shows an even more pronounced enhancement in $T_2$, whereas the 50-ppm sample exhibits only a modest field dependence.

We then compare the measured Hahn-echo decays with simulations based on a P1-only bath model using the same initial P1 concentrations and magnetic-field conditions as used experimentally (see Supplementary Note 5.1). This comparison reveals two notable discrepancies. First, although both experiment and theory show an increase of $T_2(B_0)$ with field, the P1-only model systematically overestimates $T_2$ for all three samples. Second, while the model predicts a decrease of $p(B_0)$ with increasing field, the experimental data show the opposite trend. These inconsistencies indicate that a homogeneous P1 bath—commonly assumed in prior studies[38-42]—is insufficient to account for the decoherence behavior of our samples, pointing to the presence of additional parasitic electron spins in the bath.

To address these limitations of the P1-only bath model, we focus on the *extra-defect model*, rather than the *defect-conversion model*, because it naturally explains the experimentally observed reduction of $T_2$ with increasing defect concentration [X]. From Figure 4, we determine the defect concentration [X] that reproduces the experimental $T_2$ values at $B_0$ ~ 300 G for each *V/H-related defect*. For the [$^{14}$N] = 12 ppm sample, the experimental $T_2 \approx 9.5\ \mu s$ is reproduced when adding $[V^-]$ = 4 ppm, $[VH^-]$ = 6 ppm, or $[V^+]$ = 10 ppm to the P1 bath, whereas the inclusion of a $VH^0$ defect does not reproduce the data (Figure 4b). Notably, the predicted $V^+$ concentrations are consistent with the independently inferred vacancy densities. Our CCE calculations yield $[V^+] \approx 6$ ppm for the 10-ppm sample, and $[V^+] \approx 20$ ppm for the 50-ppm sample (See Supplementary Figure 6 for the 10-ppm result), in close agreement with experimentally estimated vacancy concentrations of 8.4 ppm and 25.2 ppm, respectively (See

Method). This agreement indicates that our mixed-bath model captures the essential NV decoherence mechanisms across different bath densities.

We next examine the magnetic-field dependence of $T_2$ and $p$ in mixed baths containing P1 centers and individual V/H-related defects. The magnetic field is varied while keeping all other environmental parameters fixed, and the extracted $T_2$ and $p$ values are summarized in Figures 5d-f. Importantly, the extra-defect model reproduces the field dependence of NV coherence substantially better than the P1-only model, without introducing any adjustable parameters. We note that the 12-ppm sample shows the largest discrepancy under the P1-only model (see Supplementary Note 5.1), but this mismatch is significantly reduced when $V^+$ defects are included. For example, in Figure 5b, the calculated $T_2$ values of 7.9 μs at 33.7 G and 10.5 μs at 142 G closely match the experimental values of 7.7 μs and 8.9 μs, respectively. In contrast, the P1-only model predicts a sharp increase in $T_2$ near 100 G followed by saturation (Supplementary Figure 5b,e). The mixed bath including $V^+$ defects (orange dashed lines in Figure 5b,e) instead shows a gradual increase of $T_2$, consistent with experiment. The corresponding $p$ values remain nearly constant around 0.95 with small fluctuations, exhibiting significantly improved agreement relative to the P1-only model, with an average deviation of only ~4.5% across the field range.

A similar level of agreement is obtained for $V^-$-mixed baths (red dashed lines in Figure 5b,e), which reproduce the field-dependent trends of both $T_2$ and p. In contrast, the mixed bath including $VH^-$ defects (blue dashed lines) displays qualitatively different behavior: $T_2$ shows a more pronounced rise near 150 G, and $p$ decreases from 1.30 to 0.96 with a peak of 1.43 in the 100–200 G range. Moreover, in the 10-ppm and 50-ppm samples (Figure 5a,c,d,f), $VH^-$-mixed baths produce excessively rapid coherence decay near ~100 G, inconsistent with the experimental profiles. These comparisons indicate that single-vacancy defects ($V^+$ and $V^-$) play a dominant role in reproducing the experimentally observed field-dependent decoherence across all three samples. Overall, our results demonstrate that a mixed-bath description incorporating vacancy-related spins is essential for accurately capturing NV ensemble decoherence, whereas a single-species P1 bath model is insufficient.

**DISCUSSION**

In summary, we developed and experimentally validated a predictive computational framework for quantum decoherence in realistic defect-rich diamond environments by combining first-principles calculations with quantum many-body simulations and experiment. By analyzing homogeneous spin baths composed of individual defect species, we demonstrated that each PM defect uniquely influences NV spin coherence, with the resulting $T_2$ values following the trend $P1 > NVH^- > NV^0 \approx VH^0 > V^+$ bath. This finding casts doubts on the common approximation adopted in previous studies, where non-P1 defects are treated as generic bare electron spin baths. Instead, we find that NV coherence is strongly governed by defect-specific electronic structure properties, including hyperfine interaction strength, symmetry-axis orientation arising from geometric configurations, and electron spin quantum number.

In our study we investigated two mixed-bath models that reflect realistic defect formation processes in NV-diamond synthesis: (i) a defect-conversion model for nitrogen-related defects and (ii) an extra-defect model for vacancy/hydrogen-related defects. These two models reveal qualitatively different mechanisms by which additional PM species modify NV decoherence in a P1-dominant bath. In the defect-conversion model, we observe an enhancement of $T_2$ from 25.1 $\mu s$ to 40.4 $\mu s$ for $R_{NX} < 0.4$, indicating that compositional inhomogeneity among nitrogen-related defects can suppress flip-flop dynamics and extend coherence under experimentally relevant conditions ($R_{NX} < 0.3$)[28, 32, 33, 37]. In contrast, the introduction of V/H-related defects generally reduces $T_2$. Specifically, the presence of $V^-$ and $VH^-$ strongly reduces NV coherence, and it is largely independent of bath composition. These results demonstrate that $T_2$ in mixed baths is governed not only by the total defect concentration but also critically by the bath composition and defect identity. More generally, these results demonstrate that defect composition constitutes an essential descriptor of quantum decoherence alongside defect density.

We find that experimental results are accurately reproduced by a mixed-bath model consisting of P1 centers and vacancy-related defects, rather than by the conventionally adopted P1-only model. The extracted $V^+$ defect concentrations of approximately 6 ppm, 10 ppm, and 20 ppm for the 10-ppm, 12-ppm, and 50-ppm samples, respectively. These estimates are consistent with independently inferred vacancy concentrations of 8.4 ppm and 25.2 ppm for the 10-ppm and 50-ppm samples. Moreover, the mixed-bath model successfully reproduces the magnetic-field dependence of both $T_2$ and the stretched-exponent parameter $p$. Our analysis indicates that single-vacancy defects ($V^-$ , $V^+$ ) play an important role in determining NV coherence, while the contributions of $VH^-$ and $VH^0$ defects are comparatively minor. Furthermore, the agreement between theory and experiment demonstrates that heterogeneous defect populations can be inferred directly from coherence measurements when combined with atomistically informed spin-bath simulations.

Beyond identifying dominant decoherence pathways, the framework provides a quantitative route for evaluating how specific defect populations influence quantum coherence before materials synthesis or processing. The results indicate that both defect density and defect composition must be considered when optimizing coherence, with vacancy-related defects generally degrading coherence while controlled compositional heterogeneity among nitrogen-related defects can suppress bath dynamics and prolong coherence. These findings establish practical design principles for tailoring defect environments in diamond quantum materials.

Additionally, our results establish a basis for understanding NV ensemble decoherence beyond the P1-only picture, although further developments will be needed to describe the full complexity of real samples. In particular, the $^{13}C$ nuclear-spin bath was not included, and the real sample may contain more than two types of spin defects. Determining the dominant impurity species in the sample would require further analysis, for example by combining DEER measurements with microscopic calculations, which is beyond the scope of the present work.

Although demonstrated for diamond NV ensembles, the computational framework developed here is broadly applicable to defect-based quantum materials in which coherence is limited by complex spin environments. By integrating first-principles defect characterization with quantum many-body decoherence simulations, the approach establishes a direct connection between atomistic defect properties and experimentally observable quantum coherence. This capability enables both predictive evaluation of decoherence in realistic materials environments and microscopic identification of hidden defect populations from coherence measurements. More broadly, our work provides a general framework for understanding and predicting quantum decoherence in complex materials and highlights the importance of defect composition as a fundamental descriptor of quantum functionality.

## MATERIALS AND METHODS

### Sample preparation

Thin nitrogen doped (10 ppm ≤ [N] ≤ 50 ppm) CVD diamond layer was grown on top of an electronic grade diamond (N < 5 ppb ) crystal by Element Six Ltd. (ES, [$^{14}$N] = 12 ppm, 9-μm-thick), Applied Diamond Inc. (AD, [$^{14}$N] = 50 ppm, 40-μm-thick), and Ulm university (UU, [$^{15}$N] = 10 ppm, 0.5-μm-thick), respectively. The total nitrogen concentration in the overgrown layer is provided by the manufacturers or secondary ion mass spectroscopy (SIMS). $NV^-$ centers in ES were created by Element Six Ltd. and the concentration was approximately 3.6 ppm. The other two diamond crystals were electron irradiated to create $NV^-$ centers. The electron irradiation energy was 1 MeV and the total dosages were $1 \times 10^{19}/cm^2$ (UU) and $3 \times 10^{19}/cm^2$ (AD), respectively. They were annealed in vacuum at 800°C for 4 hours and 1000 °C for 2 hours. An $NV^-$ yield of UU and AD was approximately estimated to be 10%, derived from the intensity of photon fluorescence counts.

**Optical measurements**

NV coherence measurements were done using a home-built confocal scanning laser microscope. An acousto-optic modulator (Gooch&Housego 3200-121) allowed time-gating of a 300 mW, 532 nm diode-pumped solid-state laser (MGL-III-532-300mW). We aligned the telescope lens to focus the laser beam passing through the AOM. Double-path method was used to enhance isolation ratio, and the 1st order diffraction beam was coupled to a single-mode fiber. Reflected after a dichroic filter, fiber coupled green beam passed through an air objective (Zeiss EC Epiplan-Apochromat 100x/0.95), which then focused the beam onto a target NV center. Diamond sample was fixed on a three-axis piezo controlled stage (P-562.3CD), and a copper wire was placed on top of a diamond for microwave signal delivery. Red fluorescence signal from NV was collected back through the same objective, then onto a silicon avalanche photodetector (Perkin Elmer SPCM-ARQH-12). Then, we placed a pinhole (diameter 50 μm) with f = 150 mm telescope and removed 20 % of light due to unfocused background light. To align external bias magnetic field, we used a ½ inch cylindrical neodymium magnet, attached to 3 axis motorized stage (NRT100) for automated field alignment. Magnetic field misalignment angle was less than 1 degree. All measurements were carefully time-synchronized using pulse generating FPGA system with clock rate of 100 MHz. Finally, all measurements were performed at room temperature.

**Quantum bath model and CCE method**

Simulations of NV decoherence in the presence of PM defects were performed using the quantum bath model and the central spin model. The NV center was treated as a two-level system with effective spin states $m_s = 0$ and $m_s = -1$. A magnetic field was applied along the [111] direction, and a Hahn-echo pulse sequence ( $\pi/2$ - $\tau$ - $\pi$ - $\tau$ - $\pi/2$ ) was incorporated in the simulation. In the quantum bath model, PM defects were randomly distributed in space at a given density, with their electron and nuclear spin states. Notably, we include all degeneracies of the ground state for each PM defect. Therefore, one of the possible ground state configurations is randomly assigned to each PM defect in the spin bath. The spin interaction tensors for each PM defect were obtained from DFT-based electron spin densities. Meanwhile, other spin-spin interactions, such as the dipolar coupling between the NV center and the defect's electron spins or the HF coupling with nuclear spins, were treated using a point-dipole approximation. To handle the many-body system consisting of hundreds of PM defects, we employed the cluster-correlation expansion (CCE), specifically the CCE-2e2n method discussed in a previous study[39]. We note that the CCE-2e2n method effectively captures all relevant spin interactions between spins associated with the defect by treating electron and nuclear spins on an equal footing, thus enabling a more precise simulation of bath dynamics (See Supplementary Note 6.1). Our results were obtained by ensemble averaging over 20 distinct bath configurations, and for each configuration, we performed an average over 20 different bath states. Details of the convergence tests can be found in the Supplementary Note 6.2. Finally, we extracted the coherence time $T_2$ and $p$ values by fitting the function $L(2\tau) = \exp\left[-\left(\frac{2\tau}{T_2}\right)^p\right]$.

**Density functional theory calculations**

Spin interaction tensors for each PM defect were computed with the Vienna Ab initio Simulation Package (VASP) using the Perdew, Burke, and Ernzerhof (PBE) exchange-correlation functional along with the projector augmented-wave (PAW) pseudopotentials. A plane-wave cutoff energy (ENCUT) of 700 eV was used, and the Brillouin zone was sampled at the Γ-point for an orthorhombic supercell containing 576 atoms. Additionally, spin polarization was included in all calculations, and the resulting charge and spin densities were used to extract spin interaction tensors relevant to NV decoherence simulations.

**REFERENCES**

[1] J.R. Maze, P.L. Stanwix, J.S. Hodges, S. Hong, J.M. Taylor, P. Cappellaro, L. Jiang, M.G. Dutt, E. Togan, A. Zibrov, Nanoscale magnetic sensing with an individual electronic spin in diamond. Nature **455** 644–647 (2008).
[2] C.L. Degen, F. Reinhard, P. Cappellaro, Quantum sensing. Rev. Mod. Phys. **89** 035002 (2017).
[3] J.F. Barry, J.M. Schloss, E. Bauch, M.J. Turner, C.A. Hart, L.M. Pham, R.L. Walsworth, Sensitivity optimization for NV-diamond magnetometry. Rev. Mod. Phys. **92** 015004 (2020).
[4] N. Aslam, H. Zhou, E.K. Urbach, M.J. Turner, R.L. Walsworth, M.D. Lukin, H. Park, Quantum sensors for biomedical applications. Nature Reviews Physics **5** 157–169 (2023).
[5] D.R. Glenn, D.B. Bucher, J. Lee, M.D. Lukin, H. Park, R.L. Walsworth, High-resolution magnetic resonance spectroscopy using a solid-state spin sensor. Nature **555** 351–354 (2018).
[6] Y. Wang, W. Zhang, H. Chai, Z. Zhang, S. Lin, X. Qin, J. Du, Fully integrated quantum magnetometer based on nitrogen-vacancy centers. Phys. Rev. Appl. **23** 034008 (2025).
[7] C. Delle Donne, M. Iuliano, B. van der Vecht, G.M. Ferreira, H. Jirovská, T.J.W. van der Steenhoven, A. Dahlberg, M. Skrzypczyk, D. Fioretto, M. Teller, P. Filippov, A.R.P. Montblanch, J. Fischer, H.B. van Ommen, N. Demetriou, D. Leichtle, L. Music, H. Ollivier, I.

te Raa, W. Kozlowski, T.H. Taminiau, P. Pawełczak, T.E. Northup, R. Hanson, S. Wehner, An operating system for executing applications on quantum network nodes. Nature **639** 321–328 (2025).
[8] C.E. Bradley, S.W. de Bone, P.F.W. Möller, S. Baier, M.J. Degen, S.J.H. Loenen, H.P. Bartling, M. Markham, D.J. Twitchen, R. Hanson, D. Elkouss, T.H. Taminiau, Robust quantum-network memory based on spin qubits in isotopically engineered diamond. Npj Quantum Inf. **8** 122 (2022).
[9] M. Pompili, S.L.N. Hermans, S. Baier, H.K.C. Beukers, P.C. Humphreys, R.N. Schouten, R.F.L. Vermeulen, M.J. Tiggelman, L. dos Santos Martins, B. Dirkse, S. Wehner, R. Hanson, Realization of a multinode quantum network of remote solid-state qubits. Science **372** 259–264 (2021).
[10] H.P. Bartling, J. Yun, K.N. Schymik, M. van Riggelen, L.A. Enthoven, H.B. van Ommen, M. Babaie, F. Sebastiano, M. Markham, D.J. Twitchen, T.H. Taminiau, Universal high-fidelity quantum gates for spin qubits in diamond. Phys. Rev. Appl. **23** 034052 (2025).
[11] M.H. Abobeih, Y. Wang, J. Randall, S.J.H. Loenen, C.E. Bradley, M. Markham, D.J. Twitchen, B.M. Terhal, T.H. Taminiau, Fault-tolerant operation of a logical qubit in a diamond quantum processor. Nature **606** 884–889 (2022).
[12] C.E. Bradley, J. Randall, M.H. Abobeih, R.C. Berrevoets, M.J. Degen, M.A. Bakker, M. Markham, D.J. Twitchen, T.H. Taminiau, A Ten-Qubit Solid-State Spin Register with Quantum Memory up to One Minute. Phys. Rev. X. **9** 031045 (2019).
[13] C. Zhang, F. Shagieva, M. Widmann, M. Kübler, V. Vorobyov, P. Kapitanova, E. Nenasheva, R. Corkill, O. Rhrle, K. Nakamura, H. Sumiya, S. Onoda, J. Isoya, J. Wrachtrup, Diamond Magnetometry and Gradiometry Towards Subpicotesla dc Field Measurement. Phys. Rev. Appl. **15** 064075 (2021).
[14] T. Wolf, P. Neumann, K. Nakamura, H. Sumiya, T. Ohshima, J. Isoya, J. Wrachtrup, Subpicotesla Diamond Magnetometry. Phys. Rev. X. **5** 041001 (2015).
[15] D.M. Toyli, D.J. Christle, A. Alkauskas, B.B. Buckley, C.G. Van de Walle, D.D. Awschalom, Measurement and Control of Single Nitrogen-Vacancy Center Spins above 600 K. Phys. Rev. X. **2** 031001 (2012).
[16] G. Kucsko, P.C. Maurer, N.Y. Yao, M. Kubo, H.J. Noh, P.K. Lo, H. Park, M.D. Lukin, Nanometre-scale thermometry in a living cell. Nature **500** 54–58 (2013).
[17] S. Hsieh, P. Bhattacharyya, C. Zu, T. Mittiga, T.J. Smart, F. Machado, B. Kobrin, T.O. Höhn, N.Z. Rui, M. Kamrani, S. Chatterjee, S. Choi, M. Zaletel, V.V. Struzhkin, J.E. Moore, V.I. Levitas, R. Jeanloz, N.Y. Yao, Imaging stress and magnetism at high pressures using a nanoscale quantum sensor. Science **366** 1349–1354 (2019).
[18] B. Fortman, L. Mugica-Sanchez, N. Tischler, C. Selco, Y. Hang, K. Holczer, S. Takahashi, Electron–electron double resonance detected NMR spectroscopy using ensemble NV centers at 230 GHz and 8.3 T. Journal of Applied Physics **130** (2021).
[19] L. Rondin, J.-P. Tetienne, T. Hingant, J.-F. Roch, P. Maletinsky, V. Jacques, Magnetometry with nitrogen-vacancy defects in diamond. Rep. Prog. Phys. **77** 056503 (2014).
[20] J. Rovny, S. Gopalakrishnan, A.C.B. Jayich, P. Maletinsky, E. Demler, N.P. de Leon,

Nanoscale diamond quantum sensors for many-body physics. Nature Reviews Physics **6** 753–768 (2024).
[21] N. Aslam, M. Pfender, P. Neumann, R. Reuter, A. Zappe, F. Fávaro de Oliveira, A. Denisenko, H. Sumiya, S. Onoda, J. Isoya, J. Wrachtrup, Nanoscale nuclear magnetic resonance with chemical resolution. Science **357** 67–71 (2017).
[22] J.F. Barry, M.J. Turner, J.M. Schloss, D.R. Glenn, Y. Song, M.D. Lukin, H. Park, R.L. Walsworth, Optical magnetic detection of single-neuron action potentials using quantum defects in diamond. Proceedings of the National Academy of Sciences **113** 14133–14138 (2016).
[23] B.A. McCullian, A.M. Thabt, B.A. Gray, A.L. Melendez, M.S. Wolf, V.L. Safonov, D.V. Pelekhov, V.P. Bhallamudi, M.R. Page, P.C. Hammel, Broadband multi-magnon relaxometry using a quantum spin sensor for high frequency ferromagnetic dynamics sensing. Nat. Commun. **11** 5229 (2020).
[24] P. Bhattacharyya, W. Chen, X. Huang, S. Chatterjee, B. Huang, B. Kobrin, Y. Lyu, T.J. Smart, M. Block, E. Wang, Z. Wang, W. Wu, S. Hsieh, H. Ma, S. Mandyam, B. Chen, E. Davis, Z.M. Geballe, C. Zu, V. Struzhkin, R. Jeanloz, J.E. Moore, T. Cui, G. Galli, B.I. Halperin, C.R. Laumann, N.Y. Yao, Imaging the Meissner effect in hydride superconductors using quantum sensors. Nature **627** 73–79 (2024).
[25] J.M. Taylor, P. Cappellaro, L. Childress, L. Jiang, D. Budker, P.R. Hemmer, A. Yacoby, R. Walsworth, M.D. Lukin, High-sensitivity diamond magnetometer with nanoscale resolution. Nat. Phys. **4** 810–816 (2008).
[26] L.M. Pham, D. Le Sage, P.L. Stanwix, T.K. Yeung, D. Glenn, A. Trifonov, P. Cappellaro, P.R. Hemmer, M.D. Lukin, H. Park, A. Yacoby, R.L. Walsworth, Magnetic field imaging with nitrogen-vacancy ensembles. New J. Phys. **13** 045021 (2011).
[27] N. Bar-Gill, L.M. Pham, C. Belthangady, D. Le Sage, P. Cappellaro, J.R. Maze, M.D. Lukin, A. Yacoby, R. Walsworth, Suppression of spin-bath dynamics for improved coherence of multi-spin-qubit systems. Nat. Commun. **3** 858 (2012).
[28] C. Shinei, Y. Masuyama, M. Miyakawa, H. Abe, S. Ishii, S. Saiki, S. Onoda, T. Taniguchi, T. Ohshima, T. Teraji, Nitrogen related paramagnetic defects: Decoherence source of ensemble of NV− center. Journal of Applied Physics **132** (2022).
[29] J. Achard, V. Jacques, A. Tallaire, Chemical vapour deposition diamond single crystals with nitrogen-vacancy centres: a review of material synthesis and technology for quantum sensing applications. Journal of Physics D: Applied Physics **53** 313001 (2020).
[30] J.P. Tetienne, L.T. Hall, A.J. Healey, G.A.L. White, M.A. Sani, F. Separovic, L.C.L. Hollenberg, Prospects for nuclear spin hyperpolarization of molecular samples using nitrogen-vacancy centers in diamond. Phys. Rev. B **103** 014434 (2021).
[31] Y. Mindarava, R. Blinder, C. Laube, W. Knolle, B. Abel, C. Jentgens, J. Isoya, J. Scheuer, J. Lang, I. Schwartz, B. Naydenov, F. Jelezko, Efficient conversion of nitrogen to nitrogen-vacancy centers in diamond particles with high-temperature electron irradiation. Carbon **170** 182–190 (2020).
[32] T. Luo, L. Lindner, J. Langer, V. Cimalla, X. Vidal, F. Hahl, C. Schreyvogel, S. Onoda, S.

Ishii, T. Ohshima, D. Wang, D.A. Simpson, B.C. Johnson, M. Capelli, R. Blinder, J. Jeske, Creation of nitrogen-vacancy centers in chemical vapor deposition diamond for sensing applications. New J. Phys. **24** 033030 (2022).
[33] A.M. Edmonds, C.A. Hart, M.J. Turner, P.-O. Colard, J.M. Schloss, K.S. Olsson, R. Trubko, M.L. Markham, A. Rathmill, B. Horne-Smith, W. Lew, A. Manickam, S. Bruce, P.G. Kaup, J.C. Russo, M.J. DiMario, J.T. South, J.T. Hansen, D.J. Twitchen, R.L. Walsworth, Characterisation of CVD diamond with high concentrations of nitrogen for magnetic-field sensing applications. Mater. quantum technol. **1** 025001 (2021).
[34] P. Deák, B. Aradi, M. Kaviani, T. Frauenheim, A. Gali, Formation of NV centers in diamond: A theoretical study based on calculated transitions and migration of nitrogen and vacancy related defects. Phys. Rev. B **89** 075203 (2014).
[35] A.M. Edmonds, U.F.S. D'Haenens-Johansson, R.J. Cruddace, M.E. Newton, K.M.C. Fu, C. Santori, R.G. Beausoleil, D.J. Twitchen, M.L. Markham, Production of oriented nitrogen-vacancy color centers in synthetic diamond. Phys. Rev. B **86** 035201 (2012).
[36] C.B. Hartland, A study of point defects in CVD diamond using electron paramagnetic resonance and optical spectroscopy, University of Warwick, (2014).
[37] C. Findler, R. Blinder, K. Schüle, P. Balasubramanian, C. Osterkamp, F. Jelezko, Detecting nitrogen-vacancy-hydrogen centers on the nanoscale using nitrogen-vacancy centers in diamond. Phys. Rev. Mater. **8** 026203 (2024).
[38] E. Bauch, S. Singh, J. Lee, C.A. Hart, J.M. Schloss, M.J. Turner, J.F. Barry, L.M. Pham, N. Bar-Gill, S.F. Yelin, R.L. Walsworth, Decoherence of ensembles of nitrogen-vacancy centers in diamond. Phys. Rev. B **102** 134210 (2020).
[39] H. Park, J. Lee, S. Han, S. Oh, H. Seo, Decoherence of nitrogen-vacancy spin ensembles in a nitrogen electron-nuclear spin bath in diamond. Npj Quantum Inf. **8** 95 (2022).
[40] J.C. Marcks, M. Onizhuk, N. Delegan, Y.-X. Wang, M. Fukami, M. Watts, A.A. Clerk, F.J. Heremans, G. Galli, D.D. Awschalom, Guiding diamond spin qubit growth with computational methods. Phys. Rev. Mater. **8** 026204 (2024).
[41] P. Schätzle, R. Ghassemizadeh, D.F. Urban, T. Wellens, P. Knittel, F. Reiter, J. Jeske, W. Hahn, Spin coherence in strongly coupled spin baths in quasi-two-dimensional layers. Phys. Rev. B **110** L220302 (2024).
[42] H. Park, M. Onizhuk, E. Lee, H. Lim, J. Lee, S. Oh, G. Galli, H. Seo, Quantum decoherence of nitrogen-vacancy spin ensembles in a nitrogen spin bath in diamond under dynamical decoupling. arXiv preprint arXiv:2503.05404 DOI (2025).
[43] V. Stepanov, S. Takahashi, Determination of nitrogen spin concentration in diamond using double electron-electron resonance. Phys. Rev. B **94** 024421 (2016).
[44] S. Li, H. Zheng, Z. Peng, M. Kamiya, T. Niki, V. Stepanov, A. Jarmola, Y. Shimizu, S. Takahashi, A. Wickenbrock, Determination of local defect density in diamond by double electron-electron resonance. Phys. Rev. B **104** 094307 (2021).
[45] Z.-H. Wang, S. Takahashi, Spin decoherence and electron spin bath noise of a nitrogen-vacancy center in diamond. Phys. Rev. B **87** 115122 (2013).
[46] S. Takahashi, R. Hanson, J. Van Tol, M.S. Sherwin, D.D. Awschalom, Quenching spin

decoherence in diamond through spin bath polarization. Phys. Rev. Lett. **101** 047601 (2008).
[47] G. De Lange, T. Van Der Sar, M. Blok, Z.-H. Wang, V. Dobrovitski, R. Hanson, Controlling the quantum dynamics of a mesoscopic spin bath in diamond. Sci. Rep. **2** 382 (2012).
[48] G. de Lange, Z.H. Wang, D. Ristè, V.V. Dobrovitski, R. Hanson, Universal Dynamical Decoupling of a Single Solid-State Spin from a Spin Bath. Science **330** 60–63 (2010).
[49] T. Teraji, T. Taniguchi, S. Koizumi, Y. Koide, J. Isoya, Effective Use of Source Gas for Diamond Growth with Isotopic Enrichment. Applied Physics Express **6** 055601 (2013).
[50] C. Pellet-Mary, P. Huillery, M. Perdriat, A. Tallaire, G. Hétet, Optical detection of paramagnetic defects in diamond grown by chemical vapor deposition. Phys. Rev. B **103** L100411 (2021).
[51] K.C. Wong, S.L. Ng, K.O. Ho, Y. Shen, J. Wu, K.T. Lai, M.Y. Leung, W.K. Leung, D.B.R. Dasari, A. Denisenko, J. Wrachtrup, S. Yang, Microscopic Study of Optically Stable Coherent Color Centers in Diamond Generated by High-Temperature Annealing. Phys. Rev. Appl. **18** 024044 (2022).
[52] G. Wang, C. Li, H. Tang, B. Li, F. Madonini, F.F. Alsallom, W.K. Calvin Sun, P. Peng, F. Villa, J. Li, P. Cappellaro, Manipulating solid-state spin concentration through charge transport. Proceedings of the National Academy of Sciences **120** e2305621120 (2023).
[53] L.B. Hughes, Z. Zhang, C. Jin, S.A. Meynell, B. Ye, W. Wu, Z. Wang, E.J. Davis, T.E. Mates, N.Y. Yao, K. Mukherjee, A.C. Bleszynski Jayich, Two-dimensional spin systems in PECVD-grown diamond with tunable density and long coherence for enhanced quantum sensing and simulation. APL Materials **11** (2023).
[54] X. Yu, E.J. Villafranca, S. Wang, J.C. Jones, M. Xie, J. Nagura, I. Chi-Durán, N. Delegan, A.B. Martinson, M.E. Flatté, D.R. Candido, G. Galli, P.C. Maurer, Engineering Dark Spin-Free Diamond Interfaces. arXiv preprint arXiv:2504.08883 DOI (2025).
[55] J. Steeds, S. Kohn, Annealing of electron radiation damage in a wide range of Ib and IIa diamond samples. Diamond and related materials **50** 110–122 (2014).
[56] N. Nunn, S. Milikisiyants, E.O. Danilov, M.D. Torelli, L. Dei Cas, A. Zaitsev, O. Shenderova, A.I. Smirnov, A.I. Shames, Electron irradiation-induced paramagnetic and fluorescent defects in type Ib high pressure–high temperature microcrystalline diamonds and their evolution upon annealing. Journal of Applied Physics **132** (2022).
[57] J. Isoya, H. Kanda, Y. Uchida, S. Lawson, S. Yamasaki, H. Itoh, Y. Morita, EPR identification of the negatively charged vacancy in diamond. Phys. Rev. B **45** 1436 (1992).
[58] V.V. Dobrovitski, A.E. Feiguin, R. Hanson, D.D. Awschalom, Decay of Rabi Oscillations by Dipolar-Coupled Dynamical Spin Environments. Phys. Rev. Lett. **102** 237601 (2009).
[59] R. de Sousa, Electron spin as a spectrometer of nuclear-spin noise and other fluctuations, Electron spin resonance and related phenomena in low-dimensional structures, Springer2009, pp. 183–220.
[60] E.I. Rosenthal, C.P. Anderson, H.C. Kleidermacher, A.J. Stein, H. Lee, J. Grzesik, G. Scuri, A.E. Rugar, D. Riedel, S. Aghaeimeibodi, G.H. Ahn, K. Van Gasse, J. Vučković, Microwave Spin Control of a Tin-Vacancy Qubit in Diamond. Phys. Rev. X. **13** 031022 (2023).

[61] W. Yang, W.-L. Ma, R.-B. Liu, Quantum many-body theory for electron spin decoherence in nanoscale nuclear spin baths. Rep. Prog. Phys. **80** 016001 (2017).
[62] W.M. Witzel, M.S. Carroll, Ł. Cywiński, S. Das Sarma, Quantum decoherence of the central spin in a sparse system of dipolar coupled spins. Phys. Rev. B **86** 035452 (2012).
[63] W.M. Witzel, M.S. Carroll, A. Morello, Ł. Cywiński, S. Das Sarma, Electron Spin Decoherence in Isotope-Enriched Silicon. Phys. Rev. Lett. **105** 187602 (2010).
[64] W. Yang, R.-B. Liu, Quantum many-body theory of qubit decoherence in a finite-size spin bath. Phys. Rev. B **78** 085315 (2008).
[65] N. Zhao, S.-W. Ho, R.-B. Liu, Decoherence and dynamical decoupling control of nitrogen vacancy center electron spins in nuclear spin baths. Phys. Rev. B **85** 115303 (2012).
[66] M. Onizhuk, G. Galli, Decoherence of solid-state spin qubits: a computational perspective. arXiv preprint arXiv:2405.18535 DOI (2024).
[67] R. Ghassemizadeh, W. Körner, D.F. Urban, C. Elsässer, Coherence properties of NV-center ensembles in diamond coupled to an electron-spin bath. Phys. Rev. B **110** 205148 (2024).
[68] A.M. Ferrari, M. D'Amore, K.E. El-Kelany, F.S. Gentile, R. Dovesi, The NV0 defects in diamond: A quantum mechanical characterization through its vibrational and Electron Paramagnetic Resonance spectroscopies. Journal of Physics and Chemistry of Solids **160** 110304 (2022).
[69] C. Glover, M.E. Newton, P. Martineau, D.J. Twitchen, J.M. Baker, Hydrogen Incorporation in Diamond: The Nitrogen-Vacancy-Hydrogen Complex. Phys. Rev. Lett. **90** 185507 (2003).
[70] G. Di Palma, F.S. Gentile, V. Lacivita, W.C. Mackrodt, M. Causà, R. Dovesi, N2 positively charged defects in diamond. A quantum mechanical investigation of the structural, electronic, EPR and vibrational properties. Journal of Materials Chemistry C **8** 5239–5247 (2020).
[71] C.V. Peaker, First principles study of point defects in diamond, Newcastle University, (2018).
[72] C. Glover, M.E. Newton, P.M. Martineau, S. Quinn, D.J. Twitchen, Hydrogen Incorporation in Diamond: The Vacancy-Hydrogen Complex. Phys. Rev. Lett. **92** 135502 (2004).
[73] K. Czelej, M.R. Zemła, P. Śpiewak, K.J. Kurzydłowski, Quantum behavior of hydrogen-vacancy complexes in diamond. Phys. Rev. B **98** 235111 (2018).
[74] I. Takács, V. Ivády, Accurate hyperfine tensors for solid state quantum applications: case of the NV center in diamond. Communications Physics **7** 178 (2024).
[75] P.G. Baranov, H.J. Von Bardeleben, F. Jelezko, J. Wrachtrup, Magnetic Resonance of Semiconductors and Their Nanostructures, Springer (2017).

## Acknowledgements

**Funding:** This work was supported by the National Research Foundation (NRF) of Korea grant

funded by the Korean government (MSIT) (No. 2023R1A2C1006270, RS-2025-25454922, RS-2025-02219034), by Creation of the Quantum Information Science R&D Ecosystem (Grant No. RS-2023-NR068116), and by the education and training program of the Quantum Information Research Support Center (No. 2021M3H3A103657313), funded through the NRF of Korea funded by the Korea government (MSIT). This work was supported by the Institute of Information & Communications Technology Planning & Evaluation (IITP) grant funded by the Korea government (MSIT) (No. 2022-0-01026, RS-2025-25464252). This work was supported by the National Supercomputing Center with supercomputing resources, including technical support (KSC-2025-CRE-0553). This material was based upon work supported by, or in part by, the KIST institutional program (Project No. 26E0011). FJ acknowledge support of Carl Zeiss Foundation (QPhoton and Ultrasens-Vir), EU via project FLORIN, BMFTR via project EXTRASENS and QSENS and ERC via HyperQ Synergy Grant.

**Competing interests:** The authors declare they have no competing interests.

**Data, code, and materials availability:** The code for CCE calculations can be found at https://github.com/HuijinPark/CCEX. All data needed to evaluate and reproduce the results in the paper are present in the paper and/or the Supplementary Materials. This study did not generate any new materials